\documentstyle[mlpp,epsf]{article}
\lefthead{GR\"UN AND LANDGRAF}
\righthead{COLLISIONAL CONSEQUENCES}
\authoraddr{E. Gr\"un, 
MPI f. Kernphysik, 
D-69117 Heidelberg,
Germany,
(Eberhard.Gruen@mpi-hd.mpg.de)}
\slugcomment{To appear in the {\it Journal of Geophysical
Research} (JGR special section: ``Interstellar Dust and the
Heliosphere'')}
\begin{document}
\title{Collisional Consequences of Big Interstellar Grains}
\author{Eberhard Gr\"un}
\affil{Max-Planck-Institut f\"ur Kernphysik, Heidelberg, Germany}
\author{Markus Landgraf}
\affil{NASA Johnson Space Center, Houston, Texas, U.S.A.}
\begin{abstract}
Identification by the Ulysses spacecraft of interstellar grains inside
the planetary system provides a new window for the study of diffuse
interstellar matter. Dust particles observed by Ulysses and confirmed
by Galileo are more massive ($\geq 10^{-13}\;{\rm g}$) than the
'classical' interstellar grains. Even bigger grains ($\approx
10^{-7}\;{\rm g}$) were observed in form of interstellar meteors. We
analyze the consequences of the plentiful existence of massive grains
in the diffuse interstellar medium. Astronomically observed
'classical' interstellar grains can be described by a size
distribution ranging from about 5 to 250 nm in radius (about
$10^{-18}$ to $10^{-13}\;{\rm g}$). Lifetimes of these particles, due
to mutual collisions in interstellar space, can be as short as $10^{5}
f$ years, where $f = 10$ to $1000$, is the fraction of total lifetime to
the time when grains are exposed to supernova shocks. Shattering is a
source of the smallest of these grains, but grains more massive than
about $10^{-16}\;{\rm g}$ of the classical interstellar grain population are
rapidly destroyed. When applying the same shattering mechanism to the
more massive grains found recently, we suggest that collisions of
particles bigger than about $10^{-15}\;{\rm g}$ provide a source for smaller
grains. Because massive grains couple to the interstellar gas only
over large (100 to 1000 pc) length scales, the cosmic abundance ratio
of gas-to-dust needs only to be preserved averaged over corresponding
volumes of space.
\end{abstract}
\twocolumn
\section{Introduction}

Information about interstellar dust has been extremely
limited. Interstellar dust is usually astronomically observed by the
obscuration (extinction) of starlight. Extinction of starlight,
especially in the UV, provides a means to quantify the dust cross
section along the line-of-sight between the observer and a distant
star (e.g. {\it Mathis}, 1990). On a large scale, the starlight extinction
is typically observable as averaged over galactic (kpc) distances. On
a local scale, the amount of interstellar dust is indirectly inferred
in the following way: the abundances of elements heavier than helium
measured along the line-of-sight to a nearby star are compared with
``cosmic abundances'' and any missing heavy elements are assumed to be
tied up in grains, which gives the amount of interstellar dust
(cf. {\it Frisch}, 1981; {\it Frisch et al.}, 1999).

Galactic dust is believed to originate from a variety of different
stars and stellar phenomena: e.g. carbon-rich stars, red giants, or
supernovae, all of which provide dust with a different and
characteristic chemical and isotopic signature (for a review, see
{\it Dorschner and Henning}, 1995). A variety of presolar grains have been
identified in primitive meteorites: e.g. diamonds, graphite, silicon
carbide, or corundum grains ({\it Zinner}, 1998). However, the identified
grains are only a minute fraction of the total material that went into
the protoplanetary disk. The composition of the bulk of the grains is
largely unknown.

Interstellar dust grains passing through the planetary system have
been detected by the dust detector onboard the Ulysses spacecraft
({\it Gr\"un et al.}, 1993). These observations provided unique
identification of interstellar grains by three characteristics: 1. At
Jupiter's distance, the grains have been found to move on retrograde
trajectories, opposite to the orbits of most interplanetary grains,
and this flow direction coincides with that of interstellar gas ({\it
Witte et al.}, 1993), 2. a constant flux has been observed at all
latitudes above the ecliptic plane ({\it Gr\"un et al.}, 1997; {\it
Kr\"uger et al.}, 1999), while interplanetary dust displayed a strong
concentration towards the ecliptic plane, and 3. the measured speeds
(despite their substantial uncertainties) of interstellar grains were
higher than the locel Solar System escape speed, even if one neglects
radiation pressure effects ({\it Gr\"un et al.}, 1994).

It is known that neutral interstellar gas flows through the solar
system with a speed of $26\;{\rm km}\;{\rm s}^{-1}$ from the direction
of $253^\circ$ ecliptic longitude and $5^\circ$ ecliptic latitude
({\it Lallement}, 1993, {\it Witte et al.}, 1993). The flow of
interstellar dust detected by Ulysses coincided with this direction
({\it Frisch et al.}, 1999), and persisted during Ulysses' tour over
the poles of the sun. Interstellar dust was identified as close as 1.8
AU from the sun at ecliptic latitudes above $50^\circ$ ({\it Gr\"un et
al.}, 1997). In addition, dust impact measurements from the Hiten
satellite, in high eccentric orbit about the Earth, gave indications
that interstellar meteoroids indeed reached the Earth's orbit ({\it
Svedhem et al.}, 1996). Other dust measurements by Galileo confirmed
these findings ({\it Baguhl et al.}, 1995; {\it Baguhl et al.},
1996). Modeling of the Galileo and Ulysses data suggested that up to
$30\%$ of dust flux, with masses above $10^{-13}\;{\rm g}$, at 1 AU is
of interstellar origin ({\it Gr\"un et al.}, 1997; {\it Landgraf},
1999).

The masses of clearly identified interstellar grains range from
$10^{-15}\;{\rm g}$ to above $10^{-11}\;{\rm g}$ with a maximum at about
$10^{-13}\;{\rm g}$. More recently, even bigger ($\approx 10^{-7}\;{\rm g}$)
interstellar meteors have been reliably identified ({\it Taylor et
al.}, 1996, {\it Baggaley}, 1999). The deficiency of small grain
masses ($\leq 10^{-15}\;{\rm g}$) is not solely introduced by the
detection threshold of the Galileo and Ulysses instruments, but
indicates a depletion of small interstellar grains in the
heliosphere. Estimates of the filtering of $0.1\;{\rm \mu m}$-sized
and smaller electrically charged grains in the heliospheric bow shock
region and in the heliosphere itself ({\it Frisch et al.}, 1999; {\it
Linde and Gombosi}, 1999; {\it Landgraf}, 1999), show that these
particles are strongly impeded from entering the inner planetary
system by the interaction with the ambient magnetic field that is
carried by the solar wind. We will assume in the following analysis
that, in addition to the big particles observed within the
heliosphere, there exist also the astronomically observed smaller
grains in the local interstellar medium.

Astronomically observed interstellar dust is conveniently described by
the MRN size distribution ({\it Mathis, et al.}, 1977) ranging
from 5 to 250 nm in size (about $10^{-18}$ to $10^{-13}\;{\rm g}$). The
size distribution of these particles can be described by a power law
distribution (with an exponent of -3.5) and the total amount is
related to the volatile density in the local cloud. On the basis of
this size distribution, {\it Jones et al.} (1996) find that collisional
shattering is a source of the smallest of these grains, but grains
bigger than about 30 nm are rapidly depleted, much faster than they
can be replenished by condensation in stellar outflows ({\it Jones et
al.}, 1994). In this study we investigate whether collisions with the
newly found bigger particles could provide a source for grains bigger
than those generated by collisions from the classic interstellar
grains alone (cf. {\it Jones et al.}, 1997).

The mass distribution of interstellar grains measured by Galileo and
Ulysses only overlaps with the biggest masses of the 'classical' MRN
distribution, and extends to much bigger particles. For this analysis,
we will assume that the mass distribution of the big particles extends
to the mass of the observed interstellar radar meteors ($\approx
10^{-7}\;{\rm g}$, see {\it Landgraf et al.}, 1999, for a discussion of the
mass distribution observed by spacecraft and radar
observations). While the grain flux up to $10^{-10}\;{\rm g}$ is well
characterized by spacecraft measurements the meteor flux at about
$3\times 10^{-7}\;{\rm g}$ has still large uncertainties. {\it Landgraf et
al.} (1999) give a cumulative mass flux with a slope of $-1.1\pm 0.1$
(corresponding to a $-4.3\pm 0.3$ slope of the size distribution) for
the Ulysses and Galileo measurements alone, however, the radar data
suggest a significantly flatter slope of the combined mass
distribution, and a slope of -4 is compatible with both
measurements. These big particles are mostly hidden from astronomical
observations because their contribution to the total cross section for
extinction is low.

Where do big grains come from? Since big grains are difficult to
observe, here we can only speculate. Stellar outflows (star dust) and
explosion shells display only small grains that are injected into the
interstellar medium. Circumstantial evidence for micron sized and
bigger grains outside of our own solar system exists mostly in dust
disks around Vega type stars (cf. eg. {\it Backman and Paresc}, 1993). It is
in molecular clouds, star forming regions and, especially, in
proto-planetary disks where one would expect grain growth, however,
the mechanism by which these large grains escape into the diffuse
medium remains unclear. Excess emission at millimeter wavelengths from
very cold dust in the Galactic plane is interpreted by {\it Rowan-Robinson}
(1992) to be due to particles of several $10\;{\rm \mu m}$ in size.

In the main part of this paper, we will demonstrate that the mere
existence of a significant number of big particles in parts of the
diffuse interstellar medium has locally profound consequences for the
evolution of interstellar material. The big particles' masses provide
a significant collisional reservoir for smaller particles, in those
interstellar regions where big particles are abundant. However, it is
not the scope of this paper to develop an evolutionary model of
interstellar dust in the local interstellar medium, especially, since
source terms for interstellar grains are not considered.

We will make simplifying assumptions for the size distribution and the
collision speed in order to arrive at an estimate of the collisional
effects of big interstellar grains. First, we assume a size
distribution of big grains that is compatible with that observed by
Ulysses, and combine it with a standard model of locally unobserved
small classic grains. The consequence of this assumption will be that
the results obtained are strictly valid only in regions of space where
such a grain size distribution is found. Furthermore, we will assume a
constant speed of $100\;{\rm km}\;{\rm s}^{-1}$, as the effective
collision speed at which particles of different sizes collide. This
speed is derived from the average speed that supernova shocks travel
through the interstellar medium (cf. {\it Jones et al.}, 1996). The
collisional effects are mitigated by the factor f that is the fraction
of total lifetime when grains are exposed to shocks, where $f = 10$ to
$1000$. While the absolute numbers may be uncertain by a factor 10,
the qualitative results that we emphasize are still valid.

The mass added by the big grains to the local interstellar medium is
not violating well founded cosmic abundance arguments. In the
discussions portion of the paper, we will argue that the dust-to-gas
mass ratio may vary locally, and that the cosmic abundance needs only
to be preserved from averages over large volumes of space.

\section{Size and mass distributions of grains in the local
interstellar medium}

The MRN size distribution of grains in the diffuse interstellar medium
is given by a power law $dn = A n_H a^{-\alpha}\!da$, with slope
$\alpha = 3.5$, and $A = 7.76\times 10^{-26}\;{\rm cm}^{-2.5}\;{\rm\
per\ H\ nucleus}$ ({\it Mathis et al.}, 1977). In the following, we will
assume $n_H = 0.3\;{\rm cm}^{-3}$, a typical value for the local
interstellar medium ({\it Frisch et al.}, 1999). The grain sizes, a,
range from $5$ to $250\;{\rm nm}$ ($m = 1.7\times 10^{-18}\;{\rm g}$ to
$2.2\times 10^{-13}\;{\rm g}$). In our discussion, we follow a similar
discussion by {\it Gr\"un et al.} (1985) for the distribution of
interplanetary dust but we use, instead of the size distribution, the
mass distribution after assuming spherical particles of density $\rho
= 3\;{\rm g}\;{\rm cm}^{-3}$. For mass distributions covering a wide
mass range, it is convenient to use the logarithmic differential
distribution:
\begin{eqnarray}
dn & = & C_{\rm MRN}m^{\frac{1 - \alpha}{3}}d\!\log m
\end{eqnarray}
with the constant
\begin{eqnarray}
C_{\rm MRN} & = & A n_H \left( \frac{3}{4 \pi \rho}
\right)^{\frac{1 - \alpha}{3}} \frac{1}{3} \ln 10.\nonumber
\end{eqnarray}

\begin{figure*}
\epsfxsize=.9\hsize
\epsfbox{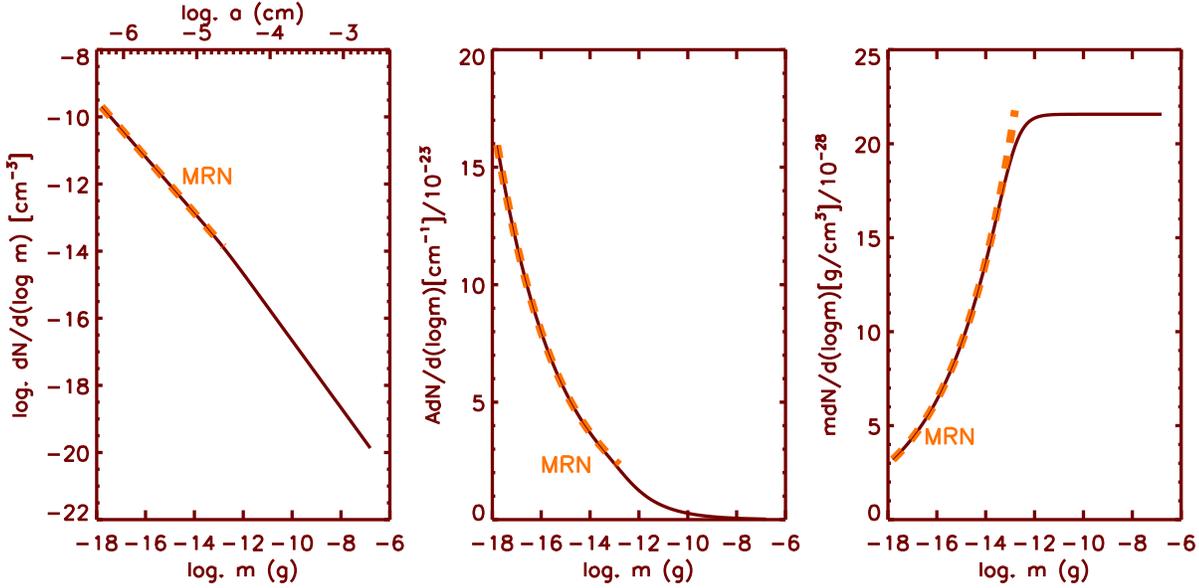}
\caption{
\it
Mass distribution and moments of the mass distribution
(logarithmic differential number-, cross section-, and mass-density
distributions) for the MRN (bold dashed) and the extended
distributions.}
\end{figure*}

The new extended mass distribution consists of three parts: a MRN-type
distribution (slope $-\alpha = -3.5$) for small particles (about
$10^{-18}\;{\rm g}$ to $10^{-14}\;{\rm g}$), a distribution with a
steeper slope ($-\alpha = -4.0$) for bigger particles (about
$10^{-12}\;{\rm g}$ to $10^{-7}\;{\rm g}$), and a transition region in
between. We have assumed that the slope of the big particle
distribution holds up to the size of the radar meteors. The extended
mass distribution is represented by

\begin{eqnarray}
dn & \propto & \left[ \frac{ \left( 1 + x \right)^{\gamma - 1} }{x^\gamma} 
\right]^{\beta \delta} da
\end{eqnarray}

with $x = \left( \frac{a}{a_t} \right)^{1/\delta}$, where $-\beta\delta
= -3.5$ is the slope for small particles and $-\beta = -4$ is the slope for
big particles, and $\delta = 0.5$ controls the transition. Again, we
transform this extended distribution to the logarithmic differential
distribution:

\begin{eqnarray}
dn & = & C_{\rm ext} \left[ \frac{\left( 1 + x \right)^{\gamma -
1}}{x^\gamma} \right]^{\beta\delta} \left( \frac{3}{4\pi\rho}
\right)^{\frac{1}{3}}\nonumber\\
&& \times \frac{1}{3} m^\frac{1}{3} d\!\log m
\end{eqnarray}

with $x = \left( \frac{m}{m_t} \right)^\frac{1}{3\delta}$, the
transition mass $m_t = 1.5 \times 10^{-13}\;{\rm g}$ and $C_{\rm ext} =
9.0\times10^{-10} n_H$. The extended distribution is valid from $m =
1.7\times 10^{-18}\;{\rm g}$
to $2.2\times 10^{-7}\;{\rm g}$ ($a = 5\;{\rm nm}$ to $a = 25\;{\rm \mu 
m}$).

The cross-section and the mass distributions are given by 
\begin{eqnarray}
\frac{dn_A}{d\!\log m} & = & A_d(m) \frac{dn}{d\!\log m}
\end{eqnarray}

with $A_d(m) = \pi^{\frac{1}{3}} \left( \frac{3m}{4\rho} \right)^{2/3}$, 
and by
\begin{eqnarray}
\frac{dn_m}{d\!\log m} & = & m \frac{dn}{d\!\log m}
\end{eqnarray}

Fig. 1 shows the MRN and extended mass distributions and their
moments: cross section density $n_A$ and mass density $n_m$
distributions. For the extended distribution, the total mass density
is $M = \int n_m d\!\log m = 1.78\times 10^{-26}\;{\rm
g}\;{\rm cm}^{-3}$ (MRN: $4.97\times 10^{-27}\;{\rm g}\;{\rm
cm}^{-3}$) and the total cross section density is $A =
\int n_A d\!\log m = 3.90\times
10^{-22}\;{\rm cm}^{2}\;{\rm cm}^{-3}$ (MRN: $3.63\times 10^{-22}\;{\rm
cm}^{2}\;{\rm cm}^{-3}$). The cross section density of the extended
distribution is not significantly increased ($+7\%$) over that of the
MRN distribution and, consequently, the interpretation of extinction
measurements at visible and UV wavelengths is not affected. However,
most mass density is contained in the big particles (the total mass of
the extended distribution is about a factor 3.5 increased over that in
the MRN distribution).

\section{Collision Dynamics}

We now determine the effects of mutual collisions in a cloud of
interstellar matter like that currently surrounding the solar
system. On an average of every few $10^{7}$ years, a supernova shock
passes through the diffuse medium with relative speeds of about 50 to
$200\;{\rm km}\;{\rm s}^{-1}$ ({\it Jones et al.}, 1996). The pressure
jump and the entrained magnetic field first accelerates the gas then
betatron acceleration causes the small grains, and eventually the
bigger grains, to reach their post shock speed. In this way,
relative speeds between grains of various sizes are introduced. In
this study, we will focus on the collisional destruction and the
generation of fragments by collisions between grains, and we will
ignore sputtering of grains by the fast moving gas, which may
additionally erode small grains in very fast shocks.  We assume a
quasi-stationary situation, by taking into account, only the shock
environment where the relative speeds between the grains are of the
order of the shock speed. Hence, we will not follow the collisional
time evolution of the size distribution, and therefore we assume that
the grain size distribution does not change with time. Taking into
account that interstellar matter cycles repeatedly through a warm and
cold interstellar medium (where it is effectively shielded from
shocks), the effective lifetimes may be a factor $f = 10$ to $1000$
times longer than calculated.

In the subsequent section, we follow the collision formulation of {\it
Gr\"un et al.} (1985) for interplanetary dust and that of {\it Jones et
al.} (1996) for interstellar dust. An important parameter describing
the effect of collisions is $\Gamma$, the ratio of the target and projectile
masses, at which a ``catastrophic'' collision occurs, i.e. a collision
that shatters the target particle completely. {\it Jones et al.}
(1996) derives this value $\Gamma = 10^{3}$ to $10^{4}$) from first
principles for interstellar particles in a shock environment. {\it
Gr\"un et al.} (1985) used values $\Gamma = 10^{5}$ to $10^{6}$) for
interplanetary collisions that were derived from laboratory
experiments with cm-sized projectiles and basalt targets at impact
speeds below $10\;{\rm km}\;{\rm s}^{-1}$. Since the {\it Jones et
al.} values were derived for the size and speed range applicable to
interstellar grains, we use their value ($\Gamma = 3\times 10^{3}$).

During the passage of a supernova shock through the interstellar
medium, the relative speeds between particles of different sizes vary
over a wide range with the maximum relative speed close to the shock
speed itself. Any target particle of a given mass is most effectively
destroyed by projectiles that are up to a factor $\Gamma$ smaller than the
target particle. Relative speeds between different massive particles
reach shock speed. Even particles of the same mass can get high
relative speeds, since shock acceleration is facilitated by the
pick-up process, where charged grains gyrate about the moving magnetic
field and hence colliding particles can be at different phases of
their gyration. Collisional effects are strongly dependent on the
collision speed (see e.g. {\it Jones et al.}, 1996), and the highest
speeds have the strongest effect. Therefore, we assume $v = 100\;{\rm
km}{\rm s}^{-1}$ as the effective speed of all catastrophic collisions
during the passage of a shock.

The collision rate $c(m_1)$ of target particles of mass m1 is given by 
\begin{eqnarray}
c(m_1) & = & -\int_{m_1/\Gamma}^{M_\infty} \sigma(m_1,m_2) v
\frac{dn(m_2)}{d\!\log m_2} d\!\log m_2\nonumber\\
\end{eqnarray}

where $\sigma(m_1,m_2)$ is the collisional cross section. 

\begin{figure}
\epsfxsize=.95\hsize
\epsfbox{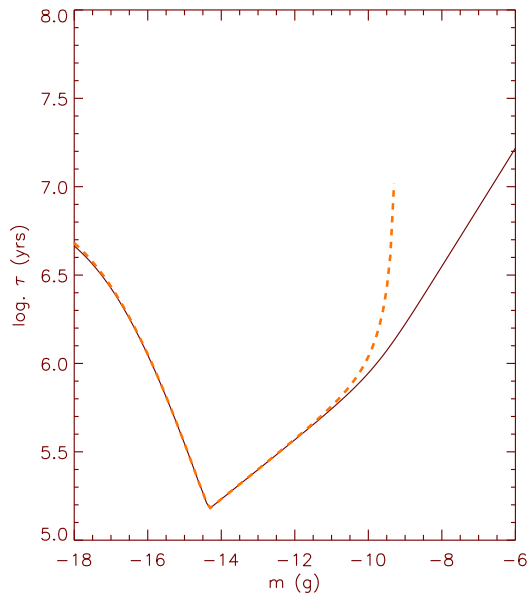}
\caption{
\it
\figurewidth{.45\hsize}
Collisional lifetimes of grains in the local interstellar
medium ($n_H = 0.3\;{\rm cm}^{-3}$) for projectiles from the MRN (bold
dashed) and the extended distributions.
\vspace{2mm}}
\end{figure}

From the collision rate we can determine the mean collisional lifetime
$\tau(m)$ of grains of mass $m$:
\begin{eqnarray}
\tau(m) & = & \frac{1}{c(m)}.
\end{eqnarray}

Fig. 2 shows that collisional lifetimes for the biggest particles of
the MRN distribution ($10^{-13}\;{\rm g}$) are of order a few $10^{5}$
years, giving rise to a fast depletion of these particles, and a
steepening slope of the MRN distribution as described by {\it Jones et
al.} (1996). The biggest particles of the extended distribution
($10^{-7}\;{\rm g}$) have lifetimes about 100 times longer than those
of the biggest MRN-particles.

In the next step, we calculate the mass density destruction rate
$d\!\dot{M}_d$ shattered in catastrophic collisions:

\begin{eqnarray}
\frac{d\dot{M}_d(m_1)}{d\!\log m_1} & = & \frac{dn(m_1)
m_1}{d\!\log m_1} v \int_{m_1/\Gamma}^{M_\infty} \sigma(m_1,
m_2)\nonumber\\
&& \times \frac{dn(m_2)}{d\!\log m_2} d\!\log m_2.
\end{eqnarray}

It must be noted that the mass of the colliding particles is not lost,
but reappears in the distribution of fragment particles. From
laboratory experiments, the fragment mass distribution is given by

\begin{eqnarray}
\frac{dG(m,m_1,m_2)}{d\!\log m} & = & \mbox{const.}\cdot m^{-\eta},\
\mbox{for}\ m<m_L \label{eqn_dg}\nonumber\\
\end{eqnarray}

with $\eta = 0.77$ (after {\it Jones et al.}, 1996), $m_L$ is the mass
of the largest fragment (we will use a value of $5\times 10^{-3}$ of
the target mass) and the constant in equation (\ref{eqn_dg}) can be evaluated
from the conservation of mass (assuming no significant losses by
evaporation).

The mass density generation rate $\dot{M}_g$ of fragments can be
calculated from

\begin{eqnarray}
\frac{d\dot{M}_g(m)}{d\!\log m} & = & m \int_{\mu}^{M_\infty}
\frac{dn(m_1)}{d\!\log m_1} v \nonumber\\
&&\times \int_{m_1/\Gamma}^{M_\infty}
\frac{dG(m,m_1,m_2)}{d\!\log m} \sigma(m_1, m_2) \nonumber\\
&&\times \frac{dn(m_2)}{d\!\log m_2} d\!\log m_2\;d\!\log m_1,
\end{eqnarray}
where $\mu = 10^{-18}\;{\rm g}$ is the mass of the smallest particle
considered as projectiles.

Fig. 3 shows the mass distributions of the destroyed particles and the
generated fragments for the extended distributions. The total mass
shattered (equals the total mass of generated fragments) per unit time
and volume for the MRN distribution is $5.7\times 10^{-4}\;{rm
g}\;{\rm m}^{-3}\;{\rm s}^{-1}$) whereas the processed mass for the
extended distribution is $1.1\times 10^{-3}\;{\rm g}\;{\rm
m}^{-3}\;{\rm s}^{-1}$). 

\begin{figure}
\epsfxsize=.95\hsize
\epsfbox{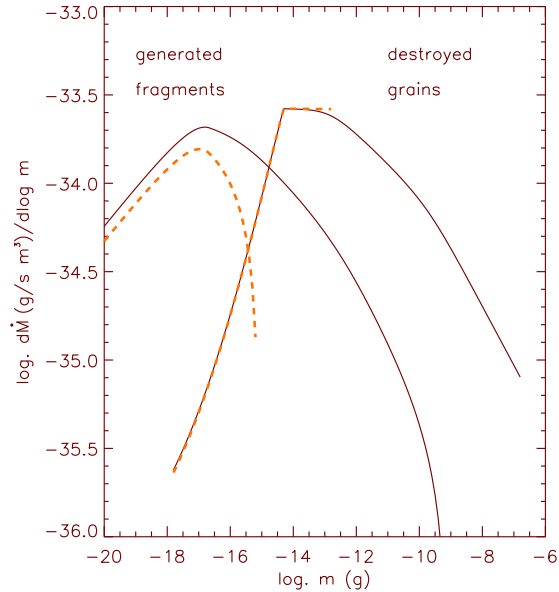}
\caption{
\it
\figurewidth{.45\hsize}
Mass distributions of destroyed particles and the generated
fragments for the MRN (bold dashed) and extended distributions.
\vspace{2mm}
}
\end{figure}

Collisions are an effective source of small
grains (cf. {\it Jones et al.}, 1996): more fragments are generated than
grains are destroyed in each mass interval below $10^{-16}\;{\rm g}$
and $10^{-15}\;{\rm g}$ for the MRN and the extended mass distribution,
respectively. The situation is reversed above these thresholds,
however, while bigger grains are rapidly lost from the MRN
distribution, the losses from the extended mass distribution are
reduced by collisional fragments that are generated up to
$10^{-10}\;{\rm g}$.

\section{Discussion}

A second consequence of the existence of big interstellar grains in
the local interstellar medium is that the gas-to-dust ratio may
deviate significantly from that derived from cosmic abundance
arguments (about $1\%$ of the total mass is in refractory dust;
cf. {\it Frisch et al.}, 1999). In this section, we show that massive dust
grains couple to the interstellar gas over much larger distance scales
than do the small classic interstellar grains. As a consequence, the
dust-to-gas mass ratio may vary locally (on scales of a few pc),
depending on the relative abundance of massive grains, and cosmic
abundance arguments are valid only averaged over large volumes of
space (100 to 1000 pc dimension). In the following section, we
calculate relevant length scales that govern grains in the diffuse
interstellar medium and discuss consequences for the gas-to-dust mass
ratio.

For comparison with other processes that affect grains in the diffuse
interstellar medium, we calculate the collisional length scale:
\begin{eqnarray}
l_{\rm coll} & = & v \tau.
\end{eqnarray}

Again we assume that dust particles move with a speed of $v =
100\;{\rm km}{\rm s}^{-1}$ with respect to the ambient medium. This
length scale is a lower limit -- it could be a factor f longer. Next we
calculate the length scales over which interstellar dust couples to
other constituents of the ambient diffuse interstellar medium: gas and
fields. The frictional scale ldrag over which the dust couples to the
gas (cf. {\it Morfill and Gr\"un}, 1979) is given by the path length it takes
for the grain to sweep-up its own mass, md, in form of interstellar
gas (hydrogen, $m_H = 1.67\times 10^{-24}\;{\rm g}$):

\begin{eqnarray}
l_{\rm drag} & = & \frac{m_d}{A_d n_H m_H}.
\end{eqnarray}

An important scale for an interstellar grain is its coupling to the
ambient magnetic field. We will assume that interstellar particles in
the local interstellar medium are charged typically to $U = 0.5 V$
({\it Gr\"un and Svestka}, 1996). Therefore, their charge is given by
$q = 4\pi\epsilon_0 U a$, with permittivity, $\epsilon_0 = 8.859\times
10^{-12}\;{\rm C}{\rm V}^{-1}\;{\rm m}^{-1}$, and $a$ in meters. The
gyro radius, $l_{\rm gyr}$, in the ambient magnetic field, $B =
0.5\;{\rm nT}$ (estimated average value in the local diffuse
interstellar medium, {\it Holzer}, 1989), is
\begin{eqnarray}
l_{\rm gyr} & = & \frac{mv}{qB}.
\end{eqnarray}

Fig. 4 compares the scales for collisional destruction, gas friction,
and gyration, assuming that the local condition are representative for
the diffuse interstellar medium over a much larger volume. The
shortest scale is the interaction with the magnetic field. If we
further assume that the interstellar magnetic field is coupled to the
gas via the ionized component, the gyro radius is the shortest scale
over which the dust couples to the gas. Nevertheless, the biggest
particles ($10^{-7}\;{\rm g}$) can travel over about one kpc before
they are captured by the magnetic field, and therefore, big particles
may be generated several 100 pc from the place where they are
found. It is evident that big particles couple over much larger scales
to the ambient diffuse interstellar medium (gas and fields) than small
particles do, therefore, they may be unrelated to the local gas
density and the small particles except via collisions.
\begin{figure}
\epsfxsize=.95\hsize
\epsfbox{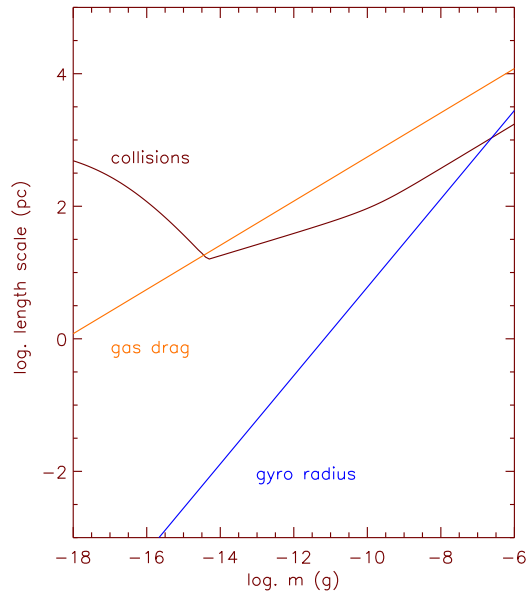}
\caption{
\it
\figurewidth{.45\hsize}
Length scales ($1\;{\rm pc} = 3\times 10^{16}\;{\rm m}$) for
collisional destruction (extended mass distribution only), gas
friction (gas density $n_H = 0.3\;{\rm cm}^{-3}$), and gyration
(surface potential $U = 0.5\;{\rm V}$, magnetic field $B = 5\;{\rm
nT}$). 
\vspace{2mm}
}
\end{figure}

Consequently
the dust-to-gas mass ratio may vary locally, and hence, the cosmic
abundance (about 1
only over large scales (100 pc to 1 kpc). Another consequence of the
much longer coupling length of big particles is that there is
everywhere a significant relative speed (of the order of $30\;{\rm
km}{\rm s}^{-1}$) between big grains and smaller grains that couple to
the local gas. Initial estimates indicate that the collisional effects
of this constant grinding may be as important for the generation of
small particles as those induced by supernova shocks.

The most important consequences of the big particle population in the
local diffuse interstellar medium are that: 1. most mass is in big
particles, 2. massive particles have long collisional lifetimes (
$10^{7}$ years), 3. big particles provide a source for small MRN-type
particles as long as big grains are present, 4. massive grains couple
to the gas over length scales of 100 to 1000 pc, 5. the dust-to-gas
mass ratio varies locally depending on the relative amount of massive
particles, 6. cosmic abundance ratio of gas to dust is only valid over
kpc distances.

The purpose of the paper is to draw the attention to effects of big
interstellar grains that have not been previously recognized. A full
evolutionary model of interstellar dust in the diffuse interstellar
medium is left to future work.

\acknowledgements
This paper was presented at the ISSI workshop on ``Interstellar Dust
and the Heliosphere'', Bern, 1998. ISSI's support for this activity is
acknowledged.

\end{document}